\documentclass[epj,twocolumn]{webofc}
\usepackage[varg]{txfonts}   % Web of Conferences font
\woctitle{The Innermost Regions of Relativistic Jets and Their Magnetic Fields}
\begin{document}
\title{First Results from {\em NuSTAR} Observations of Mkn~421}

\author{
M.~Balokovi\'{c}\inst{1}\fnsep\thanks{\email{mislavb@astro.caltech.edu}} \and
M.~Ajello\inst{2} \and R.~D.~Blandford\inst{3} \and S.~E.~Boggs\inst{2} \and F.~Borracci\inst{4} \and
J.~Chiang\inst{3} \and F.~E.~Christensen\inst{5} \and W.~W.~Craig\inst{2} \and K.~Forster\inst{1} \and 
A.~Furniss\inst{3} \and F.~F\"{u}rst\inst{1} \and G.~Ghisellini\inst{6} \and B.~Giebels\inst{7} \and
P.~Giommi\inst{8} \and B.~W.~Grefenstette\inst{1} \and C.~J.~Hailey\inst{9} \and F.~A.~Harrison\inst{1} \and
M.~Hayashida\inst{3} \and B.~Humensky\inst{9} \and Y.~Inoue\inst{3} \and
J.~E.~Koglin\inst{9} \and H.~Krawczynski\inst{10} \and
G.~M.~Madejski\inst{3} \and K.~K.~Madsen\inst{1} \and D.~L.~Meier\inst{11} \and
T.~Nelson\inst{12} \and P.~Ogle\inst{13} \and D.~Paneque\inst{4} \and
M.~Perri\inst{8,14} \and S.~Puccetti\inst{8,14} \and C.~S.~Reynolds\inst{15} \and
T.~Sbarrato\inst{6,16} \and D.~Stern\inst{11} \and G.~Tagliaferri\inst{6} \and
C.~M.~Urry\inst{17} \and A.~E.~Wehrle\inst{18} \and W.~W.~Zhang\inst{19}
} % end author list

\institute{
Cahill Center for Astronomy and Astrophysics, California Institute of Technology, Pasadena, CA 91125, USA % 1
\and
Space Sciences Laboratory, University of California, Berkeley, CA 94720, USA % 2
\and
Kavli Institute for Particle Astrophysics and Cosmology, SLAC National Accelerator Laboratory, Menlo Park, CA 94025, USA % 3
\and
Max-Planck-Institut f\"{u}r Physik, D-80805, M\"{u}nchen, Germany % 4
\and
DTU Space -- National Space Institute, Technical University of Denmark, Elektrovej 327, 2800 Lyngby, Denmark % 5
\and
INAF -- Osservatorio Astronomico di Brera, via E. Bianchi 46, I-23807 Merate, Italy % 6
\and
Laboratoire Leprince Ringuet, Ecole Polytechnique, F-91128 Palaiseau, France % 7
\and
ASI -- Science Data Center, via Galileo Galilei, I-00044 Frascati, Italy % 8
\and
Columbia Astrophysics Laboratory, Columbia University, New York, NY 10027, USA % 9
\and
Physics Department, Washington University, St. Louis, MO 63130-4899, USA % 10
\and
Jet Propulsion Laboratory, California Institute of Technology, Pasadena, CA 91109, USA % 11
\and
School of Physics and Astronomy, University of Minnesota, Minneapolis, MN 55455, USA % 12
\and
Infrared Processing and Analysis Center, California Institute of Technology, Pasadena, CA 91125, USA % 13
\and
INAF -- Osservatorio Astronomico di Roma, via Frascati 33, I-00040 Monteporzio Catone, Italy % 14
\and
Department of Astronomy, University of Maryland, College Park, MD 20742, USA % 15
\and
Dipartimento di Scienza e Alta Tecnologia, Universit\'{a} dell'Insubria, Via Valleggio 11, I-22100 Como, Italy % 16
\and
Department of Physics, Yale University, New Haven, CT 06520-8121, USA % 17
\and
Space Science Institute, Boulder, CO 80301, USA % 18
\and
NASA Goddard Space Flight Center, Greenbelt, MD 20771, USA % 19
} % end institution list

\abstract{Mkn~421 is a nearby active galactic nucleus dominated at all wavelengths by a very broad non-thermal continuum thought to arise from a relativistic jet seen at a small angle to the line of sight. Its spectral energy distribution peaks in the X-ray and TeV $\gamma$-ray bands, where the energy output is dominated by cooling of high-energy electrons in the jet. In order to study the electron distribution and its evolution, we carried out a dedicated multi-wavelength campaign, including extensive observations by the recently launched highly sensitive hard X-ray telescope {\em {NuSTAR}}, between December 2012 and May 2013. Here we present some initial results based on {\em {NuSTAR}} data from January through March 2013, as well as calibration observations conducted in 2012. Although the observations cover some of the faintest hard X-ray flux states ever observed for Mkn~421, the sensitivity is high enough to resolve intra-day spectral variability. We find that in this low state the dominant flux variations are smooth on timescales of hours, with typical intra-hour variations of $\lesssim5$\%. We do not find evidence for either a cutoff in the hard X-ray spectrum, or a rise towards a high-energy component, but rather that at low flux the spectrum assumes a power law shape with a photon index of approximately 3. The spectrum is found to harden with increasing brightness.}

\maketitle

\section{Introduction}

\label{sec:intro}

Mkn~421 is one of the nearest and best-studied blazars -- active galactic nuclei (AGN) whose relativistic jet is oriented almost directly along our line of sight. Like the other AGN of this type, Mkn~421 shows a flat radio spectrum, optical polarization, rapid and correlated variability, and other characteristics of relativistically beamed AGN. Its energy output shows the usual two peaks, located respectively in the X-ray and TeV $\gamma$-ray bands, which is typical for the high-peaked BL~Lac (HBL) class~\cite{urry+padovani-1995}. Its proximity and brightness in many spectral bands make it an important object to study in the context of AGN jet physics.

The non-thermal and polarized continuum observed in HBL objects from the radio to the X-ray suggests that this part of the spectral energy distribution (SED) is due to electron synchrotron radiation. The $\gamma$-ray part of the SED is likely due to the inverse Compton scattering by the same electrons responsible for the synchrotron radiation, and the seed photons are most likely the synchrotron photons internal to the jet. This scenario is the basis of the so-called Synchrotron Self-Compton (SSC) model, which has been successfully invoked to explain the complete SED of the BL~Lac class of objects; see e.g.~\cite{ulrich+1997,tavecchio+2010}. In the context of the SSC model, the variability in X-rays and high-energy $\gamma$-rays is expected to be high and correlated since they are produced by the same high-energy electrons. The time scales for energy loss of those electrons are very short, in agreement with the variability amplitude observed in Mkn~421 (spanning approximately two orders of magnitude) and rapid intra-day variability observed during epochs of high activity; see e.g.~\cite{tanihata+2003,fossati+2008}.

In order to provide insight into the radiative processes, the distribution of radiating particles, constraints on the particle acceleration, and thus the structure of the relativistic jet, we conducted a multi-wavelength study of Mkn~421 focused on the X-ray and TeV $\gamma$-ray bands. Coordinated simultaneous observations were carried out from December 2012 to May 2013 with the MAGIC and VERITAS ground-based Cherenkov-telescope arrays, and the {\em Swift} and {\em {NuSTAR}} orbiting X-ray observatories. The campaign was supported by coordinated (but not necessarily simultaneous) observations by ground-based optical, infrared and radio observatories, and the orbiting {\em Fermi} $\gamma$-ray observatory. In this report we present a preliminary analysis of the {\em {NuSTAR}} data from January through March 2013.

{\em {NuSTAR}} (Nuclear Spectroscopic Telescope Array) is a hard X-ray (3--79~keV) observatory launched into low Earth orbit in June 2012~\cite{harrison+2013}. It features the first focusing X-ray telescope to extend high sensitivity beyond the $\sim$10~keV cutoff shared by all currently active focusing soft X-ray telescopes. The inherently low background associated with concentrating the X-ray light enables {\em {NuSTAR}} to achieve approximately a one-hundred-fold improvement in sensitivity over the collimated or coded-mask instruments that operate, or have operated, in the same spectral range. Part of the {\em {NuSTAR}} primary mission is aimed at advancing our understanding of astrophysical jets through observations of archetypal blazars, such as Mkn~421, with unprecedented spectral and temporal resolution in the underexplored hard X-ray band.

\section{{\em {NuSTAR}} Observations}
\label{sec:observations}

\begin{figure}
\centering
\includegraphics[width=\columnwidth]{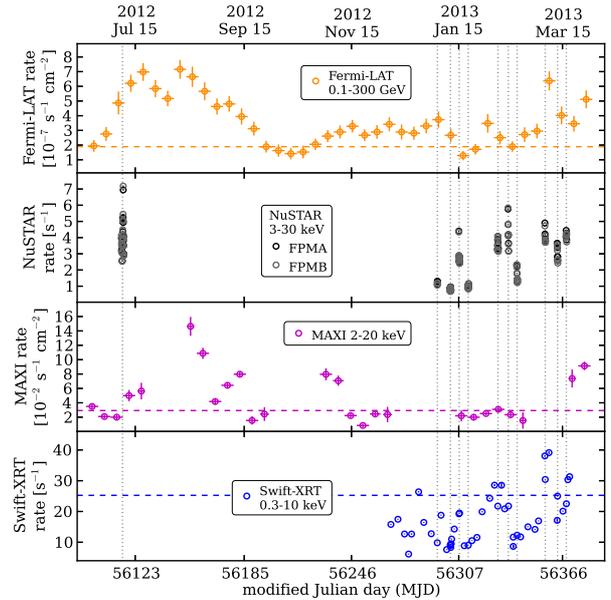}
\caption{GeV $\gamma$-ray and X-ray count rates for Mkn~421 in the period between the start of July 2012 and the end of March 2013. The {\em Fermi}-LAT, MAXI and {\em Swift}-XRT count rates were taken from publicly available data~\cite{fermi-public,maxi-public,swift-public}. Binning is weekly for the {\em Fermi}-LAT and MAXI data, per-observation for the {\em Swift}-XRT data and per-orbit for the {\em {NuSTAR}} data. Note that the uncertainties on data points from the latter two are too small to be seen in this plot. The {\em {NuSTAR}} count rates are plotted separately for its two modules, FPMA and FPMB. The dotted vertical lines mark mid-points of the {\em {NuSTAR}} observations and the dashed horizontal lines denote long-term median count rates.}
\label{fig:lc}
\end{figure}

In order to maximize the strictly simultaneous overlap of observations by {\em {NuSTAR}} and ground-based TeV $\gamma$-ray observatories during the dedicated campaign, the observation times were arranged according to visibility of Mkn~421 at the MAGIC and VERITAS sites. Coordinated observations in 2013 were performed on January 10, 15 and 20, February 6, 12 and 17, and March 5, 12 and 17. A typical {\em {NuSTAR}} observation spanned 10~hours, resulting in 15--20~ks of source exposure after accounting for orbital modulation of visibility and intervals of high background radiation. In addition to these dates, {\em {NuSTAR}} observed Mkn~421 for pointing calibration on 2012 July 7 and 8 (70~ks in total\footnote{Due to sub-optimal pointing, the count rates for this observation have increased systematic uncertainties, estimated to be $\lesssim4$\%.}) and on 2013 January 2 (10~ks). The data were reduced using the standard NuSTARDAS pipeline\footnote{\texttt{http://heasarc.gsfc.nasa.gov/docs/nustar/analysis/ nustar\_swguide.pdf}}, version~1.2.0. Figure~\ref{fig:lc} shows the observations listed above with publicly available light curves in X-ray~\cite{swift-public,maxi-public} and GeV $\gamma$-ray bands~\cite{fermi-public} in order to provide a broader context.

\section{Analysis of Flux Variability}

\label{sec:variability}

\begin{figure}
\centering
\includegraphics[width=\columnwidth]{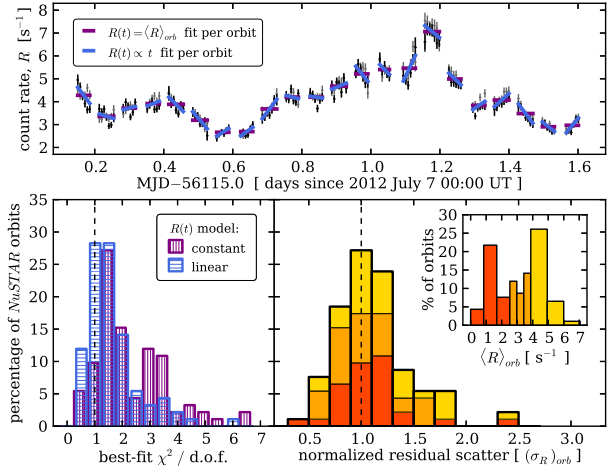}
\caption{ {\em Upper Panel:} Light curve of the July 2012 observation (FPMA in black, FPMB in grey) shown as an example for the count rate modeling; the two models fitted to each orbit of data are show in purple (constant) and blue lines (linear). {\em Lower Panels:} The colored histograms on the left-hand side (colors matching the upper panel) show the reduced $\chi^2$ distributions for the two models fitted to every {\em {NuSTAR}} orbit up to the end of March 2013. The right panel shows the distribution of the residual scatter upon subtraction of the best-fit linear trend from observed count rates in each orbit, in units of the median rate uncertainty within the orbit, ${\left( \sigma_R \right)}_{orb}$. The colors are according to the mean count rate of the orbit: the lowest-rate third in red, the mid-rate third in orange and the highest-rate third in yellow, distributed as shown in the inset. The residual scatter distribution is slightly skewed to values greater than unity, indicating that $\lesssim$20\% of orbits show excess variability on sub-orbital timescale. \label{fig:var}}
\end{figure}

The variations in count rate between the {\em {NuSTAR}} observations, and the range covered in each observation alone, are entirely dominated by the intrinsic variability of the target, i.e. they are in clear excess of the observational uncertainties. For example, the calibration observation taken in July 2012 shows flux variability of up to 30\% within an hour and a distinct increase in which the rate doubled in a roughly linear fashion over a 12-hour period (see upper panel of Figure~\ref{fig:var}). We note that it coincided with highly increased activity in the GeV~$\gamma$-ray band observed by {\em Fermi}-LAT~\cite{fermiflare-2012}, but we lack sufficient X-ray coverage to physically relate these two events. The MAXI light curve shown in Figure~\ref{fig:lc} suggests that the peak of the X-ray emission occured between mid-July and mid-August 2012, after the {\em {NuSTAR}} observation.

The campaign observations up to the end of March 2013 have covered relatively low flux states of Mkn~421, even though the lowest and the highest observed fluxes span approximately an order of magnitude. Modest count rates are apparent from both MAXI and {\em Swift}-XRT data in comparison with long-baseline light curves and the intense flaring episodes covered in the literature, e.g.~\cite{tramacere+2009}. Very high fluxes in X-ray and $\gamma$-ray bands have been observed later in the campaign, peaking in mid-April 2013~\cite{atel1,atel2,atel3}. The data from this flaring period will be presented in a dedicated publication.

With the possible exception of the lowest-flux states, the observed count rates are not consistent with being constant during any of the observations -- as demonstrated by light curves in Figures~\ref{fig:lc} and~\ref{fig:var}. In order to quantify the variability on timescales shorter than the observations, we divide the data into individual {\em {NuSTAR}} orbits, as they represent the most natural, although somewhat arbitrary, way of partitioning the data. The orbits are approximately 90~minutes long and contain up to 50~minutes of source exposure. We treat each orbit independently and fit two simple light curve models to the observed count rate in each one: a constant rate during an orbit, $R(t)={\langle R \rangle}_{orb}$, and a linear trend in time, $R(t)\propto t$. We examine the count rate in the 3--30~keV band, where the background contribution is negligible. The top panel of Figure~\ref{fig:var} provides an example of the two models fitted to the July 2012 data binned to 10-minute time intervals.

\begin{figure}
\centering
\includegraphics[width=\columnwidth]{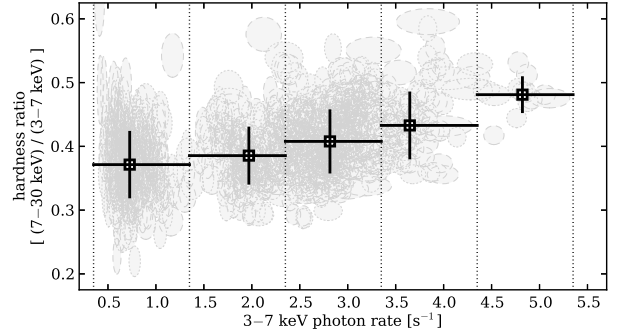}
\caption{ Hardness ratio (defined here as the ratio of count rates in the 3--7 and 7--30~keV bands) versus the 3--7~keV count rate. The grey error ellipses represent individual 10-minute data bins and their uncertainties; FPMA bins have dashed, while FPMB bins have dotted ellipse edges. The black squares and error bars (1$\sigma$ uncertainties) denote median count rate and median hardness ratio within 1~cps wide bins delimited by dotted vertical lines. The clear harder-when-brighter behaviour indicates spectral variability. \label{fig:hr}}
\end{figure}

The lower panels of Figure~\ref{fig:var} show the results of the fitting procedure performed on all orbits. We find that the majority of orbits are better described by a linear trend than a constant one. Linear trends account for most of the orbit-to-orbit variability and approximate smooth variations on super-orbital timescales of a few hours. On a 10-minute timescale, the variability amplitude typically does not exceed the observational count rate uncertainty of approximately 3\%. Based on the mildly overpopulated tail of the $\chi^2/\!$~d.o.f. distribution for the linear trend fits, we estimate that up to 20\% of orbits show excess variance beyond the simple linear trend. Subtracting the trend and comparing the residual scatter to the median rate uncertainty within each orbit, ${\left( \sigma_R \right)}_{orb}$, gives a distribution slightly skewed towards values greater than unity (see lower right panel of Figure~\ref{fig:var}). This is consistent with intrinsic sub-orbital variability on a $\sim$10-minute timescale in $\lesssim$20\% of orbits. We find no evidence for an increase in variability with flux.

\section{Spectral Analysis}

For spectral analysis we use spectra grouped to a minimum of 20 photons per bin and perform the modeling in {\em Xspec} version 12.8.0. The spectra of all {\em {NuSTAR}} observations of Mkn~421 are above the background level at least up to 25~keV and up to 40 keV in high-flux observations. The dominant background component above 25~keV is internal instrument background. With good background characterization the spectra may be used up to the high-energy end of the {\em {NuSTAR}} band at 79~keV.

We initially model the X-ray spectrum as a simple power law, $F(E)\propto E^{-\Gamma}$, where $\Gamma$ is the photon index. The model includes fixed Galactic absorption ($N_H=1.6\times10^{20}$~cm$^{-2}$) and the redshift of Mkn~421~($z=0.0308$). Fitting all full-observation spectra with a power law model and find that: i)~the mean photon index, $\langle\Gamma\rangle$, is $2.96\pm0.13$; ii)~the observations with higher mean flux systematically prefer values of $\Gamma$ lower than $\langle\Gamma\rangle$; iii)~the observations with lower mean flux systematically prefer values of $\Gamma$ higher than $\langle\Gamma\rangle$; and iv)~the best-fit $\chi^2$ increases with flux, owing to curvature apparent in the higher-flux spectra. These trends are entirely consistent with the observed hardening as the source brightens, shown in Figure~\ref{fig:hr}. The modeling implies that a $\Gamma\approx3$ power law fits the low-flux spectra well. However, the curvature and the consequent poorer fits of the power law model at high flux may either be intrinsic or simply an effect of superposition of spectra with a range of different photon indices. These two possibilities can be resolved with a more complicated spectral model and time-resolved analysis. 

\begin{figure}
\centering
\includegraphics[width=\columnwidth]{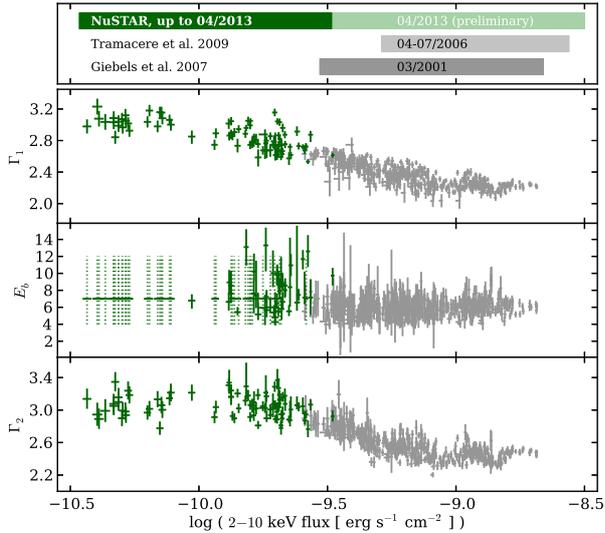}
\caption{{\em Top Panel}: Comparison of the spectral variability ranges observed with {\em {NuSTAR}} up to the end of March 2013 (dark green), and up to the end of April 2013 (light green), to the ranges previously studied two selected examples from the literature~\cite{tramacere+2009,giebels+2007}. {\em Lower Three Panels}: The green points with errorbars (1$\sigma$ uncertainties) are best-fit parameters of the broken power law model fitted to Mkn~421 spectra of individual {\em {NuSTAR}} orbits. For the orbits where the lower bound on $E_b$ is below 4~keV (i.e. close to the lower end of the {\em {NuSTAR}} band) the break energy was fixed to 7~keV and the uncertainty is estimated to range between 4 and 12~keV. The grey data points are for the same model (also 1$\sigma$ uncertainties) fitted to 2001 RXTE data, from~\cite{giebels+2007}. \label{fig:comp}}
\end{figure}

We therefore replace the power law with a purely phenomenological broken power law model: $F(E)\propto E^{-\Gamma_1}$ up to the break energy $E_b$, and $F(E)\propto E^{-\Gamma_2}$ at higher energies. By fitting spectra of each orbit individually, we partially mitigate time-averaging, as each orbit covers a relatively narrow flux range. The broken power law model provides better fits to the spectra of high-flux orbits, indicating that the curvature is intrinsic. It is, however, degenerate for low-flux orbits as both photon indices coverge to $\Gamma\approx3$ for any value of the break energy in the 4--12~keV range. In those cases we fix $E_b$ to 7~keV, which is the median best-fit value for the full-observation spectra.

We find evidence that the low-energy photon index ($\Gamma_1$) strongly varies with flux, while the flux dependence of the high-energy index ($\Gamma_2$) is weaker in the flux range of the data presented here. The break energy is difficult to constrain since for many orbits the uncertainties in the best-fit value reach below the low-energy end of the {\em {NuSTAR}} band at 3~keV, effectively causing $\Gamma_1$ to become unbound by data. In Figure~\ref{fig:comp} we show the best-fit parameters $\Gamma_1$, $E_b$ and $\Gamma_2$ as functions of 2--10~keV flux for all {\em {NuSTAR}} orbits up to the end of March 2013. The equivalent results from~\cite{giebels+2007} are plotted for comparison, demonstrating that the spectral trends found in the {\em {NuSTAR}} data smoothly continue into the previously accessible flaring regime, now covering nearly two orders of magnitude in flux.

\section{Summary and Conclusion}

The data and the analysis presented here are preliminary, but indicative of the new results uncovered by {\em {NuSTAR}} in the hard X-ray band. Its high sensitivity enabled probing the flux state significantly fainter than previously possible with comparable spectral and temporal resolution. The observed variability ranges over an order of magnitude, including instances of flux-doubling over a half-day period and occasional variability on $\sim$10-minute timescales. In this low-flux regime we find that the hard X-ray spectrum does not cut off steeply, nor show any sign of an increase towards the inverse-Compton component of the SED, but rather saturates at $\Gamma\approx3$. The spectrum hardens with an increase in flux, which can be phenomenologically described with a broken power law model: the break energy $E_b\approx7$~keV separates the soft and the hard power law slopes, both of which change with flux. This can be understood in terms of acceleration and cooling processes of radiating particles in the Mkn~421 jet, and the resulting shape of the high-energy tail of the relativistic electron energy distribution. These results, together with a broader analysis of the multi-wavelength data more physical modeling, will be presented in more detail in a paper currently in preparation.

\vspace{-10cm}

\end{document}